\documentclass[twocolumn,aps,prb,superscriptaddress,amsmath,amssymb,superscriptaddress]{revtex4-1}
\newcommand{\pagenumbaa}{1}
\usepackage{graphicx}
\usepackage{hyperref}
\usepackage{verbatim}
\usepackage{float}
\usepackage{amsmath}
\usepackage{soul}
\usepackage{xcolor}
\usepackage[utf8]{inputenc}
\usepackage[T1]{fontenc}
\usepackage{mathptmx}

\begin{document}

\title{Large-Range Frequency Tuning of a Narrow-Linewidth Quantum Emitter}

\author{Liang Zhai}
\email{liang.zhai@unibas.ch}  
\affiliation
{Department of Physics, University of Basel, 
Klingelbergstrasse 82, CH-4056 Basel, Switzerland}
\author{Matthias C. L{\"o}bl}
\affiliation
{Department of Physics, University of Basel, 
Klingelbergstrasse 82, CH-4056 Basel, Switzerland}
\author{Jan-Philipp Jahn}
\affiliation
{Department of Physics, University of Basel, 
Klingelbergstrasse 82, CH-4056 Basel, Switzerland}
\author{Yongheng Huo}
\thanks
{Present address: Shanghai Research Center for Quantum Sciences, Shanghai 201315, China \& Shanghai Branch, CAS Center for Excellence in Quantum Information and Quantum Physics, University of Science and Technology of China, Shanghai 201315, China
}
\affiliation
{Institute for Integrative Nanosciences, IFW Dresden, Helmholtzstra{\ss}e 20, 01069
Dresden, Germany}
\affiliation
{Hefei National Laboratory for Physical Sciences at the Microscale and Department of Modern Physics, University of Science and Technology of China, Hefei 230026, China}
\author{Philipp Treutlein}
\affiliation
{Department of Physics, University of Basel, 
Klingelbergstrasse 82, CH-4056 Basel, Switzerland}
\author{Oliver G. Schmidt}
\affiliation
{Institute for Integrative Nanosciences, IFW Dresden, Helmholtzstra{\ss}e 20, 01069
Dresden, Germany}
\author{Armando Rastelli}
\affiliation
{Institute of Semiconductor and Solid State Physics, Johannes Kepler Universit{\"a}t
Linz, Altenbergerstra{\ss}e 69, 4040 Linz, Austria}
\author{Richard J. Warburton}
\affiliation
{Department of Physics, University of Basel, 
Klingelbergstrasse 82, CH-4056 Basel, Switzerland}

\begin{abstract}
A hybrid system of a semiconductor quantum dot single photon source and a rubidium quantum memory represents a promising architecture for future photonic quantum repeaters. One of the key challenges lies in matching the emission frequency of quantum dots with the transition frequency of rubidium atoms while preserving the relevant emission properties. Here, we demonstrate the bidirectional frequency-tuning of the emission from a narrow-linewidth (close-to-transform-limited) quantum dot. The frequency tuning is based on a piezoelectric strain-amplification device, which can apply significant stress to thick bulk samples. The induced strain shifts the emission frequency of the quantum dot over a total range of $1.15\ \text{THz}$, about three orders of magnitude larger than its linewidth. Throughout the whole tuning process, both the spectral properties of the quantum dot and its single-photon emission characteristics are preserved. Our results show that external stress can be used as a promising tool for reversible frequency tuning of high-quality quantum dots and pave the wave towards the realisation of a quantum dot -- rubidium atoms interface for quantum networking.  

\end{abstract}

\maketitle

\setcounter{page}{\pagenumbaa}
\thispagestyle{plain}

Semiconductor quantum dots (QDs) are among the best deterministic sources of quantum light as they generate single photons\cite{peter2015,Senellart2017,JWP2019,Liu2018} and entangled photon pairs\cite{Huber2018,Liu2019,Chen2018} with high purity and indistinguishability. An additional strength of QDs is that they can be integrated into cavities and waveguides using well-established semiconductor technology\cite{peter2015,Senellart2017}. To integrate the photon sources into long-distance photonic quantum networks, it is imperative that the single photons from two remote sources have the same frequency. This condition can hardly be met in the growth process -- there is a spread of quantum dot size and composition leading to a spread in emission frequency\cite{Lobl2019,Rastelli2004}. Therefore, frequency tuning of individual quantum emitters is required. Furthermore, the photons should be indistinguishable, a condition which places a stringent condition on the emitter: the noise should be low enough that the spectral linewidths are transform limited. Promising candidates are GaAs/AlGaAs QDs\cite{LDEre}: close-to-transform-limited linewidths have been achieved\cite{zhai2020}. Moreover, the QD emission frequencies are spectrally close to the rubidium D$_{1}$ and D$_{2}$ lines\cite{Jahn2015,Zopf2019,zhai2020,Keil2017} such that a hybrid system, a QD as single photon emitter\cite{Akopian2011} and a Rb-ensemble as memory element\cite{Wolters2017}, can be conceived\cite{Matthew2013}. Implementing such a hybrid system hinges on both QD frequency-tuning and narrow QD linewidths\cite{Wolters2017}.

Different methods have been developed for frequency tuning of QDs. Applying an electric field to a QD integrated in a diode structure can adjust its emission frequency via the quantum-confined Stark effect\cite{Patel2010,Thyrrestrup2018}. Despite many successful implements on the InAs/GaAs QD platform\cite{Patel2010, Liu2018,Thyrrestrup2018}, this charge tuning method is much less investigated on GaAs/AlGaAs QDs\cite{zhai2020,Bouet2014,Langer2014}. Furthermore, when a QD is in tunnel contact to a highly-doped gate, the QD’s charge-state undergoes serval discrete jumps as the gate voltage increases, limiting the range of electric field that one can apply to a certain exciton charge-state -- and therefore the Stark shift. For InAs/GaAs QDs, the range of the Stark shift is typically below $\sim 0.1\ \text{THz}$\cite{Lobl2017,Warburton2000,Dreiser2008}; for GaAs/AlGaAs QDs, the largest range of Stark tuning reported so far is $\sim0.24\ \text{THz}$\cite{zhai2020}. Another approach is to shift the emission frequency by applying an external stress\cite{Seidl2006,Trotta2016,Huber2018,Trotta2012}. The original approach\cite{Seidl2006} applies a stress by bonding a bulk sample to a piezo-stack to which a voltage is applied at low temperature. In this scheme, application of stress does not introduce additional noise, for instance charge noise in the semiconductor, and the narrow QD linewidths of the starting material are expected to be preserved. However, the tuning range is rather small ($\simeq 0.1\ \text{THz}$\cite{Seidl2006,Jahn2015}). Much larger tuning ranges (up to $20\ \text{THz}$\cite{Yuan2018}) have been achieved on QDs in nano-membranes\cite{Kumar2011,Zhang2014,Grim2019}, or in other nanostructures\cite{Magdalena} such as nanowires\cite{Tumanov2018}. But the membranes tend to have much higher levels of charge noise than the starting material, and, therefore, QDs with broader linewidths \cite{Kumar2011}. While successively emitted photons may demonstrate high levels of indistinguishability\cite{Eva2019}, the coherence of the photons falls off as the delay increases. Also, it becomes very difficult to interfere two photons from separate QDs\cite{Reindl2017} as the charge environments are completely uncorrelated. Achieving narrow linewidths in membranes is challenging although recent progress has been made\cite{Pursley2018}. To date, there is still no successful demonstration of large-range strain tuning of GaAs QDs with narrow linewidths. For GaAs QDs, the spectral width of the ensemble of QDs is typically $\sim5\ \text{THz}$ and careful calibration of the growth can control the central frequency to less than $1\ \text{THz}$\cite{Keil2017}. In a photon memory application, it is desirable to bring a significant fraction of the QDs into resonance with the $^{87}$Rb D$_{1}$ or D$_{2}$ line\cite{Wolters2017,Jahn2018}. Therefore, a stress-based tuning range of about $1-2\ \text{THz}$ is needed. At the same time, the technique to apply stress should not induce any additional noise.

\begin{figure*}
\vspace*{-0.2cm}
\includegraphics[width=1.36\columnwidth]{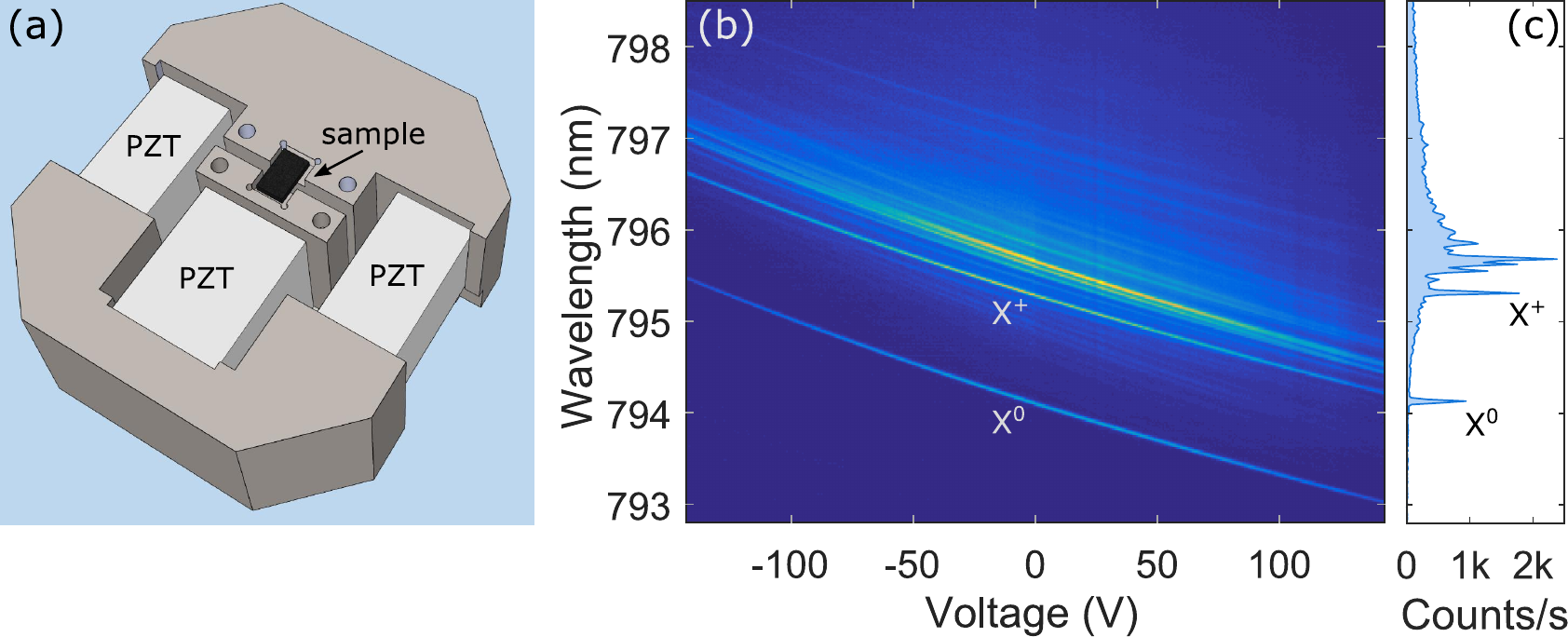}
\vspace*{-0.30cm}
\caption{(a) A sketch of the strain-amplification device. Three PZT piezoelectric stacks are mounted in parallel on a C-shaped titanium holder. The sample is glued over a narrow gap separating two movable blocks. (b) Photoluminescence (PL) from a QD around the rubidium-87 D$_{1}$ line ($794.98\ \text{nm}$). The QD can be tuned bidirectionally as a function of the voltage applied to the PZT stacks. (c) A PL spectrum of the GaAs QD at zero applied stress.
}
\label{fig:pl}
\end{figure*}

In this paper, we present a device which allows strain tuning of a QD embedded in a $100\ \mu\text{m}$-thick sample. Such a bulky sample is advantageous for minimising the charge noise since the QDs are sufficiently far away from all surfaces. We show reversible tuning of the QD's emission frequency over a range of $1.15\ \text{THz}$ -- about a thousand times more than its linewidth ($\sim1.26\ \text{GHz}$). The QD linewidth is close to the transform limit ($\sim1\ \text{GHz}$) and remains at this low level on inducing the large external strain. This frequency tuning is demonstrated on several QDs. Additionally, we show that the QD is a good single-photon emitter throughout the entire tuning range.

The GaAs/AlGaAs QD sample is fabricated by molecular-beam epitaxy on a GaAs (001) wafer using local droplet etching\cite{LDEholes,LDE2}. Unlike the commonly used Stranski-Krastanov growth-mode, the formation of droplet-etched QDs does not rely on lattice-mismatched heteroepitaxy, making the system favourable in many aspects\cite{LDEre} such as absence of the residual strain\cite{LDE2,Kuroda2013}, and ease in controlling the shape of the QDs\cite{LDE,basso2018}. After a GaAs buffer is deposited, the growth of the QD heterostructure starts with a 120 nm-thick $\text{Al}_{0.4}\text{Ga}_{0.6}\text{As}$ barrier layer, on which 0.5 monolayer (ML) of aluminium is deposited at a growth rate of 0.5 ML/s in an arsenic-depleted ambience. The Al atoms migrate and form Al-droplets on the $\text{Al}_{0.4}\text{Ga}_{0.6}\text{As}$ surface. The substrate material beneath an Al-droplet is unstable, initialising nano-hole formation\cite{LDEholes}. To facilitate this process, the sample is annealed at 600 $^\circ$C for several seconds. Then arsenic is supplied again to recrystallise the Al-rich etching residual to avoid defects. Subsequently, a 2 nm-thick GaAs layer is deposited at a rate of 0.1 ML/s. During a two-minute annealing step, diffusion into the nano-holes takes place. The filled nano-holes are finally capped with another thick $\text{Al}_{0.4}\text{Ga}_{0.6}\text{As}$ barrier to form optically active QDs.

In order to induce strain into the sample, we employ a home-built strain-amplification device following the design of Hicks {\em et al.}\cite{razor}. It is composed of three conventional lead zirconate titanate (PZT) piezoelectric stacks (Fig.~\ref{fig:pl}(a)). The device has a footprint of 24 mm $\times$ 24.5 mm, well-matched to the size of typical low temperature nano-positioners (e.g. attocube ANPx101/LT). The PZT stacks have equal lengths of $L=9$ mm and are glued onto a C-shaped titanium holder. The holder itself is fixed to the nano-positioners. The other ends of the PZT stacks connect to two movable titanium blocks, which are separated by a gap $d = 0.5\ \text{mm}$. The sample (mechanically thinned down to $100\ \mu\text{m}$ thick) straddles this gap: it is glued firmly in place. To glue the PZT stacks and the sample, we use a two-component epoxy resin adhesive (Uhu Plus Endfest 300). Upon applying a positive (negative) voltage, the central PZT stack extends (contracts) while the outer ones contract (extend), applying a compressive (tensile) stress to the sample. The three PZT stacks are connected to same electrodes, such that they all move simultaneously, minimising the shear strain. By making the cross-sectional area of the central PZT stack twice as large as the outer ones, the force applied to the sample is balanced and therefore the sample displacement in its centre is minimised. The parallel arrangement of PZT stacks also minimises the strain induced by the cooldown process\cite{razor}: when no piezo-voltage is applied, the QDs remain almost strain-free at cryogenic temperature. Owing to the fact that the PZT stacks are longer than the width of the gap, QDs in the gap region experience an amplified effective strain\cite{Yuan2018}. Assuming the sample stiffness to be much smaller than that of the PZT stacks, we expect, in the ideal case, $\epsilon_{\text{eff}} = (2L/d)\ \epsilon_{\text{PZT}}$, with $\epsilon_{\text{eff}}$ and $\epsilon_{\text{PZT}}$ denoting the effective strain in the GaAs located above the gap, and the strain of the PZT stacks, respectively. In this ideal limit, the amplification factor is $2L/d=36$. At a voltage of $+150\ \text{V}$, the unloaded PZT stack achieves $\epsilon_{\text{PZT}}=-1.20 \times 10^{-3}$ at $300\ \text{K}$. At $4\ \text{K}$, the PZT stack performance reduces by a factor of $\sim$10\cite{zhang1983}. Thus, we estimate that the device achieves $\epsilon_{\text{eff}}\sim - 4\times 10^{-3}$ on applying $+150\ \text{V}$ at $4\ \text{K}$.

We carry out photoluminescence (PL) and resonance fluorescence measurements on individual QDs with a confocal dark-field microscope \cite{dark}. The strain-amplification device is housed in a liquid helium cryostat and cooled down to $4.2\ \text{K}$. Helium gas ($25\ \text{mbar}$ at room temperature) is used as a heat exchanger between the liquid helium and the entire strain-amplification device. 

The PL measurements are performed under above-band excitation with a weak 632.8 nm He-Ne laser (intensity $\simeq 42\ \text{nW}/\mu \text{m}^{2}$). We look for QDs in the central region of the gap, where QDs experience the largest strain. The QD emission is collected by an aspheric objective lens (numerical aperture $\text{NA} = 0.71$) and is sent to a spectrometer. A typical emission spectrum of a droplet-etched GaAs/AlGaAs QD is depicted in Fig.~\ref{fig:pl}(c), where we identify two characteristic narrow lines\cite{Jahn2015,Lobl2019}: the neutral exciton (X$^{0}$) and the positively charged trion (X$^{+}$). We apply voltages of up to $\pm143\ \text{V}$ to the PZT piezo-stacks and record PL emission of the QD. As shown in Fig.~\ref{fig:pl}(b), the X$^{0}$ and X$^{+}$ lines are red-shifted (blue-shifted) in parallel by around $1.38\ \text{nm}$ ($1.08\ \text{nm}$) when experiencing tension (compression). In principle, larger stress is achievable if one further increases the voltage applied to the PZT stacks. However, we experienced issues due to electrical breakdown of the helium exchange gas when the voltage exceeded $\sim\pm 180$ V. A higher voltage (up to $\pm\ 300$ V) can potentially be applied to the PZT stacks if one reduces the pressure of the helium gas.

\begin{figure}[h!]
\vspace*{-0.3cm}
\includegraphics[width=1.01\columnwidth]{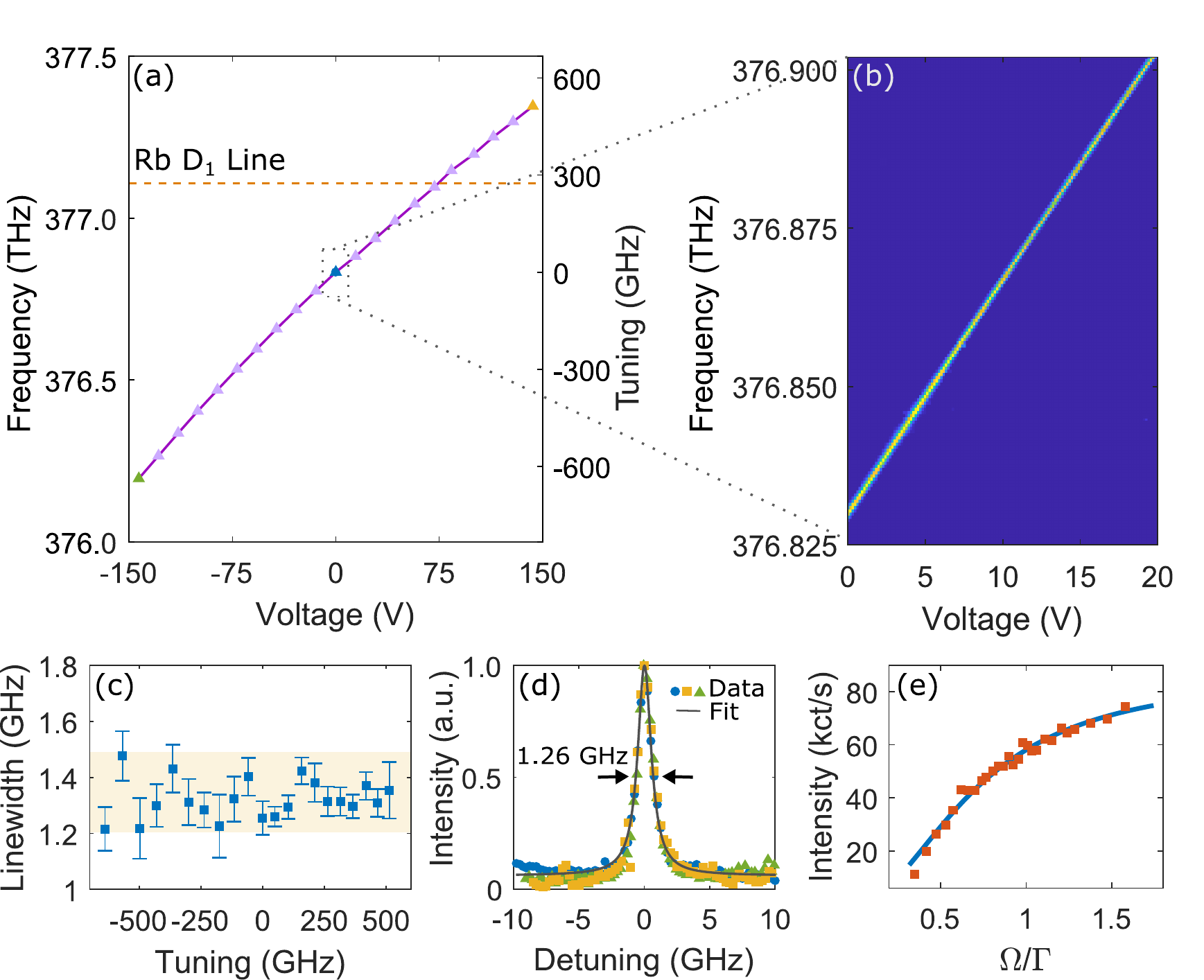}
\vspace*{-0.60cm}
\caption{(a) Frequency of single QD resonance fluorescence tuned as a function of the voltage applied to the PZT stacks (purple curve). For each data point, the frequency is determined by scanning a narrow-bandwidth laser across the QD resonance. The back-scattered laser is suppressed with a polarization-based dark-field microscope so that the QD signal can be detected. To ensure a good signal-to-background ratio, the laser suppression is monitored for every step and readjusted if necessary. The orange dashed line indicates the frequency of the $^{87}$Rb D$_{1}$ line. (b) A zoomed-in plot of a selected area in (a). The QD stays bright and narrow independent of the induced strain. The readjustment of the laser suppression is not necessary for this measurement. (c) The QD linewidth as a function of the frequency tuning with respect to the frequency at zero external stress. (d) Three different resonance fluorescence scans are plotted together in each case with normalised intensity. Blue, yellow and green symbols indicate the conditions where zero stress, maximum compression and maximum tension is applied to the QD, respectively (see (a)). A Lorentzian curve (grey curve) provides a good fit to all three data sets. The full-width-at-half-maximum (FWHM) is $1.26\pm0.02\ \text{GHz}$. (e) Intensity of the resonance fluorescence (red squares) versus normalised Rabi frequency $\Omega/\Gamma$ with a fit (blue curve) of a two-level model.}
\label{fig:rf}
\end{figure}

Resonance fluorescence measurements are performed on X$^{+}$ by scanning the frequency of a narrow-bandwidth continuous-wave (CW) laser across the QD resonance (intensity $\simeq 42\ \text{nW}/\mu \text{m}^{2}$). A very weak non-resonant laser ($\lambda = 632.8\ \text{nm}$, intensity $<0.8\ \text{nW}/\mu \text{m}^{2}$) illuminates the QD constantly during the scan, helping to stabilise the charge environment\cite{Jahn2015,Nguyen2012}. From a Lorentzian fit to the measured fluorescence intensity we determine the frequency and the linewidth of the X$^{+}$ (Fig.~\ref{fig:rf}(d)). The resonance frequency can be tuned bidirectionally as a function of strain. By applying $\pm 143\ \text{V}$ voltage to the piezo-stack, the X$^{+}$ frequency is shifted over a total range of $\Delta f = 1.15\ \text{THz}$ ($4.8\ \text{meV}$) (Fig.~\ref{fig:rf}(a)). The frequency shift per volt ($\sim 4\ \text{GHz/V}$) is 20-times larger than with the original method\cite{Seidl2006,Jahn2015} ($\sim 0.2\ \text{GHz/V}$), in which the sample is glued directly onto a PZT piezo-stack. This amplification is slightly smaller than the factor $2L/d=36$. With our device, it is possible to address the $^{87}\text{Rb}$ D$_{1}$ transition (orange dashed line in Fig.~\ref{fig:rf}(a)) with a QD which is originally far off-resonance. The strain tuning in Fig.~\ref{fig:rf}(a) shows a small non-linearity \cite{Yuan2018}: the shift of QD frequency is larger at $- 143\ \text{V}$ (tension) compared to at $+143\ \text{V}$ (compression). The origin of the non-linearity is unknown. It may be a property of the PZT stacks at $4\ \text{K}$.

The tuning of the QD emission energy $\Delta E$ with strain $\boldsymbol{\epsilon}$ can be calculated from the Pikus-Bir Hamiltonian\cite{Testelin2009}. We assume that the valence state has pure heavy-hole character and that the strain-dependence of the QD emission energy follows the strain-dependence of the GaAs band gap. In the simple case of a stress applied along the [100] direction, $\Delta E = (a_{c}-a_{v})\text{tr}(\boldsymbol{\epsilon})+\frac{b}{2}(2\epsilon_{zz}-\epsilon_{xx}-\epsilon_{yy})$, where $a_{c}$, $a_{v}$ stand for the hydrostatic deformation potentials of the conduction band and valence band, respectively, and $b$ represents the shear deformation potential\cite{Testelin2009}. For a uniaxial stress along the [100] direction\cite{Yuan2018}, $\epsilon_{xx}$, $\epsilon_{yy}$, and $\epsilon_{zz}$ are connected through the Poisson effect: $\epsilon_{yy} = \epsilon_{zz} = -\nu\epsilon_{xx}$ where $\nu$ is the Poisson ratio. For GaAs\cite{VandeWalle}, $a_{c} = -7.17\ \text{eV}$, $a_{v} = 1.16\ \text{eV}$,  $b = -1.7\ \text{eV}$ and $\nu = 0.318$. This results in $\Delta E/\epsilon_{xx}=-1.91\ \text{eV}$. In practice, the stress is applied along the [110] direction\cite{Seidl2006}. In this case, we find that $\Delta E/\epsilon_{110}=-1.80\ \text{eV}$ ($-435\ \text{THz}$). (The slight difference with respect to stress along [100] arises from the anisotropy in the stiffness tensor for GaAs.) Based on the measured frequency shift and this calculation, we estimate the maximum strain experienced by the QD at $+ 143\ \text{V}$ ( $- 143\ \text{V}$) to be $\epsilon_{110} = - 1.18 \times 10^{-3} ( + 1.47 \times 10^{-3})$. The heavy hole-light hole mixing in GaAs QDs could be slightly affected\cite{Yuan2018} by the applied stress, potentially reducing the $\epsilon_{110}$ values.

Figure.~\ref{fig:rf}(b) shows the resonance fluorescence of a selected area in Fig.~\ref{fig:rf}(a) with a higher resolution. The QD resonance features a bright and narrow line as its frequency is tuned continuously by the strain. The linewidth ($\gamma$) of the QD, shown as a function of the frequency in Fig.~\ref{fig:rf}(c), stays narrow throughout the whole tuning process. $\gamma$ fluctuates slightly with a standard deviation of 0.07 GHz and a mean value of 1.33 GHz, but there is no obvious dependence on the transition frequency. Figure.~\ref{fig:rf}(d) plots three different resonance fluorescence scans, where different colours denote different applied stress (see Fig.~\ref{fig:rf}(a)). The grey curve is a Lorentzian fit to the blue data points (zero strain), indicating a linewidth of $1.26\pm0.02\ \text{GHz}$. The same fit matches the other two data-sets similarly well. Despite the sizeable shift in emission energy, the spectral properties (such as spectral shape and linewidth) of the QD emission remain unaffected by the strain. We describe the tuning with a dynamic factor, the ratio of the total tuning range ($R$) to the resonance fluorescence linewidth, and find $R/\gamma \simeq 920$. We observed similar results for three other QDs in the region of the sample that is close to the middle of the gap.

\begin{figure}
\vspace*{-0.0cm}
\includegraphics[width=1.0\columnwidth]{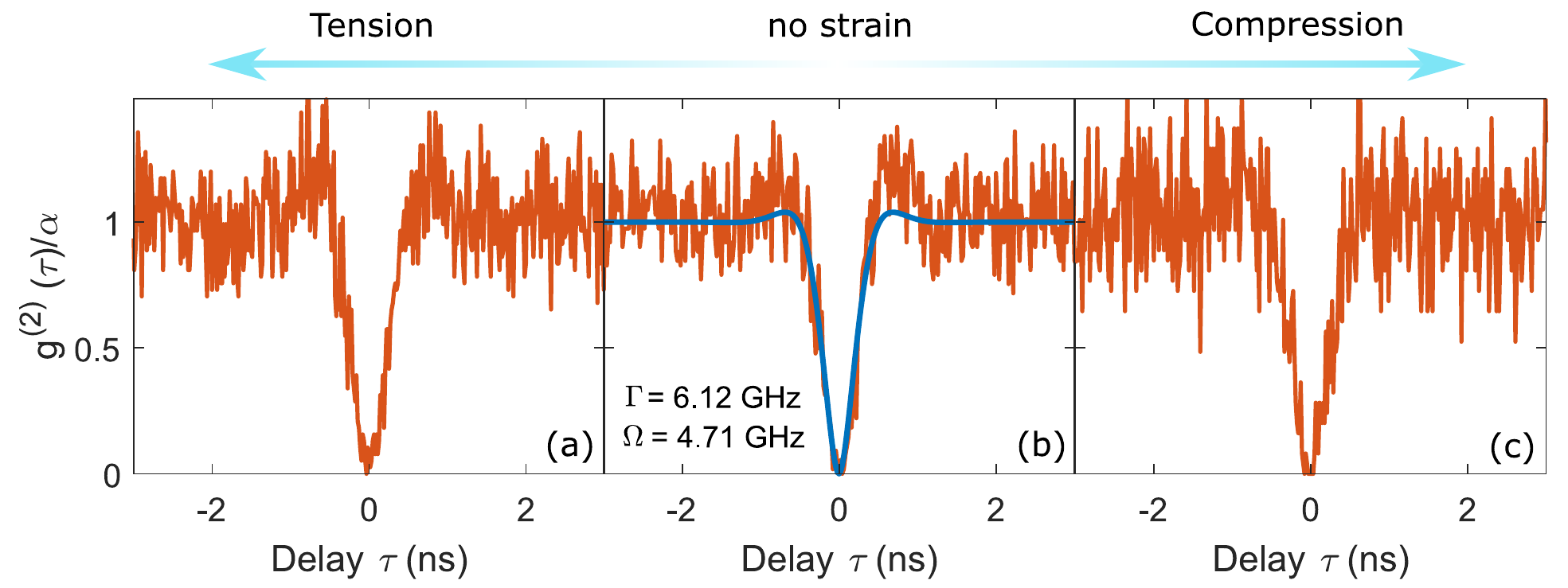}
\vspace*{-0.5cm}
\caption{Measured second-order correlation function $g^{(2)}(\tau)/\alpha$ at (a) maximum tension, (b) zero stress, (c) maximum compression under resonant CW excitation. In all three cases, the dip around zero delay ($\tau = 0$) drops close to zero, indicating single-photon emission from the QD. The fitting function Eq.~(1) is implemented to subplot (b), yielding a decay rate $\Gamma = 2\pi\times0.97\ \text{GHz}$ and a Rabi frequency $\Omega = 4.71\ \text{GHz}$. The Rabi frequency might be slightly different in (a) and (c) since the QD position relative to the laser beam might change slightly when experiencing the stress. The measured second-order correlation function is normalised to one at times of a few nanoseconds. Similar to observations in Ref. \ \onlinecite{Jahn2015}, the $g^{(2)}$ shows a blinking on longer timescales. The blinking is a result of noise and is indicated by the factor $\alpha$ ($\alpha\ \sim\ 10$). To normalise the $g^{(2)}$ correctly, the method in Ref.\ \onlinecite{Lobl2020} is used.
}
\label{fig:g2}
\end{figure}

In order to characterise the single-photon property of the emission, we perform a Hanbury Brown and Twiss (HBT) measurement under CW resonant excitation. We send the QD signal through a 50:50 beamsplitter and on to two superconducting nanowire single-photon detectors. Figure.~\ref{fig:g2} displays the measured second-order correlation function $g^{(2)}(\tau)$ as a function of delay $\tau$ in three different conditions: (a) with maximum tension ($-143\ \text{V}$ applied), (b) zero stress ($0\ \text{V}$), (c) with maximum compression ($+143\ \text{V}$ applied). In all three cases, the QD photons are time-tagged for over an hour with stable count rates (drifts < 3\% within 60 minutes), demonstrating that our strain tuning technique is suitable for measurements involving long integrations over time. When changing the applied stress, such a stability is reached after a waiting time on the minute time-scale.

Within a two-level model, $g^{(2)}(\tau)$ is given by\cite{Jahn2015}:
\begin{equation}
g^{(2)}(\tau) = 1-e^{-\frac{3\Gamma\tau}{4}}\left(\textnormal{cos}\lambda\tau +\frac{3\Gamma}{4\lambda}\textnormal{sin}\lambda\tau\right)
\label{eqn:1}
\end{equation}
with $\lambda = \sqrt{\Omega^{2}-\frac{\Gamma^{2}}{16}}$, where $\Gamma$ represents the upper-level decay-rate, and $\Omega$ the Rabi frequency. The model neglects the effect of spectral wandering. Spectral wandering, which can be effectively considered as a randomly varying detuning $\delta$, has an impact on $g^{(2)}(\tau)$ since a non-zero $\delta$ increases the effective Rabi frequency \cite{zhai2020,Loudon2000,Rezai2019}. Assuming the spectral wandering follows a probability distribution $P(\delta)$, and taking into account that the signal intensity is reduced when the excitation laser is detuned from the QD resonance, the second-order correlation function becomes: $\tilde{g}^{(2)}(\tau)\propto\int  d\delta\ g^{(2)}(\tau, \delta) P(\delta)L(\delta)^2$, where $g^{(2)}(\tau, \delta)$ is the second-order correlation function for a certain detuning $\delta$, and $L(\delta)$ is the Lorentzian lineshape of the QD resonance fluorescence depanding on the values of $\Gamma$, $\Omega$ and without spectral wandering. In our case, we find that the second-order correlation function is only weakly affected by spectral wandering. The reason is that the contribution of $g^{(2)}(\tau, \delta)$ drops quadratically as the signal intensity $L(\delta)$ decreases ($\delta$ increases). 

The relationship between $\Gamma$ and $\Omega$ is extracted from the power saturation curve in Fig.~\ref{fig:rf}(e). Subsequently, based on Eq.\ \ref{eqn:1} the two-level model for $g^{(2)}(\tau)$ is fitted to the experimental data Fig.~\ref{fig:g2}(b), where we obtain $\Gamma = 6.12\ \text{GHz}$ (radiative lifetime $163\ \text{ps}$; linewidth transform-limit $\Gamma/2\pi = 0.97\ \text{GHz}$) and $\Omega = 4.71\ \text{GHz}$. The two-level model describes the data well. In particular, $g^{(2)}(0)$ is zero in the model and this crucial feature -- it signifies photon antibunching -- is observed in all the three experimental data sets (the measured $g^{(2)}(0)$ is zero to within the error of 2\%). This shows that the QD is a good single-photon emitter at zero strain and retains this property under both tensile and compressive strain.

In summary, we present bidirectional frequency tuning of a narrow-linewidth QD by external stress. Our results show that strain tuning can be used as a non-destructive method to modify QD emission frequency over large ranges, neither broadening the  linewidth nor reducing the single-photon purity. Compared to other platforms such as $\text{Pb}$($\text{Mg}_{1/3}\text{Nb}_{2/3}$)$\text{O}_{3}$-$\text{PbTiO}_{3}$ (PMN-PT) based devices, which have been widely used for nano-membranes, our strain tuning apparatus provides a convenient way of applying stress to thick wafers with long-time stability (in hours) and high precision. The tuning range can be extended further by applying larger voltages to the PZT stacks (e.g.\ $\pm 300\ \text{V}$ with reduced helium pressure), by reducing the gap in the strain device, and possibly by softening the sample. Our technique can be also applied to other materials, e.g.\ vacancy centres in diamond, for inducing strain and frequency matching\cite{Sohn2018,Teissier2014}. The next step towards the realisation of a QD-rubidium hybrid quantum network node is to improve further the QD sample quality -- to generate single-photons without blinking and with linewidths at the transform limit. This can be achieved by embedding the QDs into a suitable diode structure\cite{zhai2020}. In this case, the strain-amplification device can be used for large-range frequency tuning, while the quantum-confined Stark effect can be used as a fast switch to bring a QD in and out of resonance with rubidium atoms.

%\section{Acknowledgement}
%\label{sec:contrib}
The authors thank Arne Ludwig and Julian Ritzmann for fruitful discussions. The authors also thank Sascha Martin and the mechanical workshop at University of Basel for their help on device fabrication. L.Z. has received funding from the European Union Horizon 2020 Research and Innovation programme under the Marie Sk{\l}odowska Curie grant agreement No.~721394 (4PHOTON). L.Z., M.C.L., J.P.J., P.T. and R.J.W. acknowledge financial support from NCCR QSIT. M.C.L., J.P.J., and R.J.W. acknowledge financial support from SNSF Project No.~200020\_156637. Y.H. is supported by NSFC (No.~11774326), National Key R$\&$D Program of China (No.~2017YFA0304301) and Shanghai Municipal Science and Technology Major Project (No.~2019SHZDZX01). A.R. acknowledges support from the FWF P29603, the Linz Institute of Technology (LIT) and the LIT Lab for secure and correct systems, supported by the State of Upper Austria. 

%\section{Author Contributions}
%\label{sec:contrib}
L.Z., M.C.L., and J.P.J. carried out the experiments. Y.H., O.G.S., and A.R. grew the sample. L.Z., J.P.J., and R.J.W. designed and fabricated the device. L.Z., M.C.L., J.P.J., and R.J.W. analysed the data. R.J.W. and P.T. initiated the project. L.Z., M.C.L., and R.J.W. wrote the manuscript with inputs from all the authors.

\section*{Data Availability}
The data that support the findings of this work are available from the corresponding author upon reasonable request.
%%%%%%%%%%%%%%%%%%%%%%%%%%%%%%%%%%%%%%%%%%%%%%%%%%%%%%%%%%%%%%%%%%%%
\bibliography{refs}
\end{document}